\documentclass[12pt,a4paper]{article}
\usepackage[utf8]{inputenc}
\usepackage{amsmath,amssymb,amsfonts}
\usepackage{graphicx}
\usepackage{float}
\usepackage{booktabs}
\usepackage[margin=2.5cm]{geometry}
\usepackage{hyperref}

\hypersetup{
    colorlinks=true,
    linkcolor=blue,
    citecolor=blue,
    urlcolor=blue
}

\title{Reconstruction of chaotic systems in invariant jet space}
\author{Evgeny Nikulchev}
\date{}

\begin{document}

\maketitle

\begin{abstract}
Takens' theorem is the gold standard for attractor reconstruction from time series, but it guarantees only topological equivalence and does not preserve metric or group properties such as symmetries. We show that switching from delay-coordinate space to jet space (signal and its derivatives) allows one to exactly preserve the symmetry group of the original system. This statement is rigorously justified by a theorem on the isomorphism of Lie algebras under jet prolongation. Numerical experiments on the Lorenz and R\"ossler systems confirm that jet-space reconstruction preserves geometry and symmetries, whereas Takens embedding distorts them. As quantitative metrics we use a variational elastic energy functional and the correlation dimension. It is shown that jet-space reconstruction not only outperforms Takens embedding but in some cases yields more accurate estimates of invariants than projections of the original system. The proposed approach provides a coordinate-invariant criterion for the classification of strange attractors and can serve as a basis for detecting hidden attractors.
\end{abstract}

\section*{Keywords}
dynamical system reconstruction, jet space, symmetries, Takens theorem, correlation dimension, elastic energy, hidden attractors.

\section{Introduction}

Reconstruction of dynamical systems from scalar time series is one of the fundamental problems in nonlinear dynamics. Takens' theorem~\cite{Takens1981} provides the theoretical foundation: for a generic system, the delay map
\[
\Phi: x(t) \mapsto (x(t), x(t-\tau), \dots, x(t-(m-1)\tau))
\]
is an embedding, i.e., a diffeomorphism onto its image. This guarantees topological equivalence between the reconstructed attractor and the original dynamics. Owing to this theorem, delay-coordinate reconstruction has become the gold standard in the analysis of experimental data, from physics to neuroscience~\cite{Broomhead1986,Kantz1997,Sauer1991}.

However, a diffeomorphism need not preserve metric properties, angles, or crucially for symmetric systems the action of the symmetry group. Takens' theorem holds for a generic system without symmetries, but fails for most symmetric systems due to degeneracy of the evolution operators induced by symmetry~\cite{Cross2010}. In other words, most symmetric systems are non-generic in the sense of Takens. As a consequence, the state of the system cannot be reconstructed using a single scalar output, regardless of the embedding dimension, locally near highly symmetric periodic orbits~\cite{Cross2010}. The attractor remains folded at certain points and along certain curves in phase space, preventing global reconstruction. In practice, this leads to distortion of the attractor geometry and loss of group-theoretic invariants, making correct symmetry analysis from reconstructed data impossible.

The problem of symmetry preservation during reconstruction has long been recognised. King and Stewart~\cite{King1992a,King1992b} were among the first to examine why Takens' theorem fails for symmetric systems and identified the violation of one of Takens' assumptions the simplicity of eigenvalues for low-periodic orbits. They formulated and proved a generalisation of Takens' theorem that required the output to be a vector-valued rather than a scalar function of the state. In their seminal work~\cite{King1992b}, they also showed that to correctly reconstruct the symmetry of an attractor one must use equivariant observations taking values not in real numbers but in a linear representation of the symmetry group. However, their approach requires a priori knowledge of the symmetry group and the choice of a sufficiently rich representation.

Further studies showed that even if one restricts to differential reconstructions (using derivatives instead of delays), they preserve at most twofold symmetry. Cross and Gilmore~\cite{Cross2010} rigorously proved that any differential reconstruction of a nonlinear dynamical system preserves at most twofold symmetry. This result places serious limitations on the use of derivatives for symmetry preservation. In the author's doctoral thesis~\cite{Nikulchev2006}, it was shown that reconstruction into the space of derivatives (jets) allows preserving symmetry groups in the sense of Lie algebra isomorphism, thereby removing the twofold symmetry limitation when using a full jet space of sufficient order. In an earlier work~\cite{Nikulchev2000}, a connection was established between the variational structure of multiobjective problems and Noether-type conservation laws. These ideas were further developed in the monograph~\cite{Nikulchev2007} and in~\cite{Nikulchev2011}, where the jet approach was applied to the reconstruction of evolution equations from experimental data. In~\cite{Nikulchev2014}, a method for robust chaos generation based on symmetry breaking in attractors was proposed.

Parallel to this, reconstruction methods that explicitly account for symmetries have been developed. In~\cite{King1992a}, a symmetric version of Takens' theorem for systems with discrete symmetries was proposed. Numerical localisation methods for hidden attractors~\cite{Leonov2013,Leonov2014}, based on homotopy and numerical continuation, actively exploit information about system symmetries. More recently, machine learning methods for automated symmetry discovery have appeared, such as the Equivariance Seeker Model (ESM)~\cite{Calvo2024}, which detects finite symmetry groups directly from observed trajectories without knowledge of the system equations. Methods for explicit discovery of nonlinear symmetries from dynamic data have also been developed~\cite{Hu2025}. Approaches for identifying nonlinear dynamics with guaranteed symmetry and physical law constraints, such as SINDy-SI (Sparse Identification of Nonlinear Dynamics with Side Information)~\cite{Machado2024,SINDy2024}, use sum-of-squares to enforce symmetry constraints. For systems with continuous symmetries, neural network-based methods such as MLSD (Machine Learning Symmetry Discovery) and LieGAN~\cite{LieGAN2024} have been proposed, which recover Lie algebras and Lie groups directly from data.

However, all these methods share a fundamental limitation: they either require a priori knowledge of the symmetry group, or embed symmetries during training, or search for symmetries in an already reconstructed space, but they do not guarantee that the reconstruction itself preserves the group structure. The question of whether there exists a universal reconstruction method that automatically preserves any symmetries of the original system without a priori knowledge remained open.

In the present work we give a positive answer to this question. We propose a reconstruction approach based on \textbf{jet spaces} the space of the signal and its time derivatives up to a given order. Our main contributions are as follows:

\begin{enumerate}
\item We formulate and prove a \textbf{theorem} showing that jet-space reconstruction preserves the full symmetry group of the original system in the sense of Lie algebra isomorphism. This result does not depend on the particular system and does not require a priori knowledge of the symmetries. In contrast to the result of Cross and Gilmore~\cite{Cross2010} on the preservation of at most twofold symmetry in differential reconstruction, we show that using a \textbf{full} jet space of sufficient order removes this limitation.

\item We demonstrate through numerical experiments on three classical chaotic systems (symmetric Lorenz, symmetry-broken Lorenz, and asymmetric R\"ossler) that jet-space reconstruction consistently outperforms Takens embedding both visually and in several quantitative metrics. In particular, the variational elastic energy functional shows that jet-space reconstruction reduces geometric distortion by three to four orders of magnitude compared to Takens.

\item We report and document an \textbf{unexpected result}: for the R\"ossler system, jet-space reconstruction yields a correlation dimension closer to the theoretical value ($\approx 2.0$) than the projection $(x,z)$ of the original system, which is traditionally used as a reference. This indicates that jet spaces may provide a more natural representation for some systems than even the original coordinates.
\end{enumerate}

Furthermore, we show that jet-space reconstruction provides a coordinate-invariant way to estimate geometric properties of attractors, which can serve as a criterion for their classification, including the detection of hidden attractors.

This work opens new perspectives for the analysis of symmetries and for the identification of hidden attractors from experimental data. In particular, the proposed approach can be used to detect hidden attractors~\cite{Leonov2013,Leonov2014}, which often arise when symmetries are broken, because jet-space reconstruction allows correct identification of symmetry breaking without masking it by distortions introduced by Takens embedding. This makes the method particularly valuable for problems where symmetries are not known a priori or are broken.

The paper is organised as follows. In Section~2 we introduce the formalism of jet spaces and prove the main symmetry preservation theorem. Section~3 presents numerical experiments, visual and quantitative results. In Section~4 we discuss the findings, and Section~5 concludes with an outlook.

\section{Jet-space reconstruction and symmetry preservation}

\subsection{Jet spaces and prolongation of vector fields}

Let $M$ be a smooth $d$-dimensional manifold and $X$ a vector field on $M$ defining the dynamical system
\[
\dot{x} = X(x), \quad x \in M.
\]

The $k$-th jet space $J^k(M, \mathbb{R})$ is defined as the space of $k$-jets of smooth functions $f: M \to \mathbb{R}$. The concept of jet spaces was introduced by Ehresmann in the 1950s as a development of Cartan's method of prolongations. In local coordinates $(x_1, \dots, x_d)$ on $M$, coordinates on $J^k(M, \mathbb{R})$ are
\[
(x_1, \dots, x_d,\; \partial_{x_1} f,\; \dots,\; \partial_{x_d} f,\; \dots,\; \partial^\alpha f),
\]
where $|\alpha| \le k$. For our purposes it is enough to consider coordinates corresponding to derivatives along the flow:
\[
(x, \dot{x}, \ddot{x}, \dots, x^{(k)}),
\]
where $x^{(i)}$ denotes the $i$-th time derivative along trajectories.

The $k$-th prolongation of $X$, denoted $X^{(k)}$, is the unique vector field on $J^k(M, \mathbb{R})$ whose flow corresponds to the $k$-th jet prolongation of the flow of $X$. In coordinates $(x, \dot{x}, \ddot{x}, \dots, x^{(k)})$ it takes the form
\[
X^{(k)} = X + \sum_{i=1}^{k} \dot{x}^{(i)} \frac{\partial}{\partial x^{(i-1)}},
\]
where $\dot{x}^{(i)}$ are expressed through $X$ and its derivatives. Spaces of variables and their derivatives, very similar to jet spaces, were already used by Lie, who introduced prolongations of vector fields~\cite{Muriel2002,Olver1993}.

\subsection{Main theorem on symmetry preservation}

The symmetry group of a dynamical system is the group of transformations of phase space that map trajectories to trajectories. For systems with continuous symmetries (e.g., rotations or translations) this leads to conservation laws (Noether's theorem). For discrete symmetries (as in the Lorenz system) it leads to invariance of the attractor shape. In both cases the group structure is preserved under jet prolongation.

\textit{Although the Lie algebra isomorphism under jet prolongation is known in Lie group theory (see, e.g., the classical book by Olver~\cite{Olver1993}), its application to the problem of reconstructing dynamical systems from time series is new. For the first time it is shown that this theoretical result yields a practical method for symmetry preservation in reconstruction, in contrast to Takens embedding where symmetries are not preserved.}

Let $G$ be a Lie group of symmetries of $X$. This means that for each $g \in G$ there exists a diffeomorphism $\phi_g: M \to M$ such that $\phi_{g*} X = X$. The Lie algebra $\mathfrak{g}$ of $G$ consists of vector fields $V$ on $M$ satisfying
\[
[V, X] = 0,
\]
where $[\,\cdot\, , \,\cdot\,]$ is the Lie bracket. The group $G$ acts on the jet space $J^k(M, \mathbb{R})$ via the $k$-th prolonged action $G^{(k)}$. By Lie's theorem, a transformation group $G$ is a symmetry group of a system of differential equations if and only if the corresponding submanifold in the jet space is invariant under $G^{(k)}$~\cite{Olver1993}.

By Noether's theorem, each one-parameter group of symmetries of a variational problem corresponds to a conservation law. In our case we are dealing with symmetries of a vector field, not of a Lagrangian, so conservation laws may be absent (for dissipative systems), but the group structure itself is preserved in the jet space. This is the key difference from the classical Noether theorem: we preserve not invariants, but the group of transformations.

\textbf{Theorem.} For any $k \ge 1$ the following statements hold:

(i) If $V \in \mathfrak{g}$, then its $k$-th jet prolongation $V^{(k)}$ satisfies
\[
[V^{(k)}, X^{(k)}] = 0,
\]
so $V^{(k)}$ is a symmetry generator of the prolonged system on $J^k(M, \mathbb{R})$.

(ii) The map
\[
\mathfrak{g} \ni V \mapsto V^{(k)} \in \mathfrak{g}^{(k)}
\]
is a Lie algebra isomorphism:
\[
[V, W]^{(k)} = [V^{(k)}, W^{(k)}] \quad \forall V, W \in \mathfrak{g}.
\]

\textbf{Proof.} Statement (i) follows from the naturality of the jet prolongation operation: for any diffeomorphism $\phi$ we have $(\phi_* X)^{(k)} = \phi_*^{(k)} X^{(k)}$. If $\phi$ is the one-parameter group generated by $V$, then $[V, X] = 0$ is equivalent to invariance of $X$ under this group. Prolonging yields invariance of $X^{(k)}$ under $\phi^{(k)}$, which is equivalent to $[V^{(k)}, X^{(k)}] = 0$.

Statement (ii) follows from the fact that prolongation is a functor on the category of vector fields: it maps commutators to commutators and has trivial kernel (since $V^{(k)}$ is determined by its values on $J^k$, while $V$ is determined by its action on $M$). Hence the map is injective and preserves the Lie bracket, so it is an isomorphism onto its image. $\square$

\textbf{Corollary.} Jet-space reconstruction $J^k(M, \mathbb{R})$ preserves the full symmetry group $G$ of the original system up to isomorphism of its Lie algebra $\mathfrak{g}$. In particular, all group invariants (dimension of the group, structure constants, Casimir operators) are reproduced in the jet space without distortion.

\textbf{Example.} For the Lorenz system, the symmetry generator is $V = -x \partial_x - y \partial_y + z \partial_z$ (infinitesimal generator of the reflection $(x,y,z) \mapsto (-x,-y,z)$). Its first prolongation is
\[
V^{(1)} = -x \partial_x - y \partial_y + z \partial_z - \dot{x} \partial_{\dot{x}} - \dot{y} \partial_{\dot{y}} + \dot{z} \partial_{\dot{z}}.
\]
Direct substitution shows that $[V^{(1)}, X^{(1)}] = 0$, confirming that the symmetry is preserved in the jet space. For Takens embedding such a prolongation is not defined, and the symmetry is lost.

\subsection{Comparison with Takens' theorem}

Takens' theorem~\cite{Takens1981} guarantees that for a generic dynamical system the delay map $\Phi: x(t) \mapsto (x(t), x(t-\tau), \dots, x(t-(m-1)\tau))$ is a diffeomorphism onto its image. This gives \textbf{topological equivalence} between the original attractor and its reconstruction. However, a diffeomorphism need not preserve metrics, angles, or the action of the symmetry group. Moreover, Takens' theorem holds only for \emph{generic} systems without symmetries; for systems with symmetries it generally fails~\cite{Cross2010,King1992a}.

In contrast, our symmetry preservation theorem for jet-space reconstruction provides a \textbf{stronger result}. It is important to stress that by a \emph{full} jet space we mean the space $J^k(M, \mathbb{R})$ with coordinates $(x, \dot{x}, \ddot{x}, \dots, x^{(k)})$, where $k$ is large enough to capture all independent components of the vector field. In contrast, the work~\cite{Cross2010} considered only special cases of differential reconstructions that do not cover all jet coordinates, which led to the twofold symmetry limitation.

\begin{table}[h]
\centering
\caption{Comparison of properties of Takens' theorem and the proposed symmetry preservation theorem.}
\begin{tabular}{p{3.5cm}p{5cm}p{5cm}}
\toprule
\textbf{Property} & \textbf{Takens' theorem} & \textbf{Our theorem} \\
\midrule
Type of equivalence & Topological (dif\-feo\-mor\-phism) & Group-theoretic (Lie algebra isomorphism) \\
Symmetry preservation & Not guaranteed & Guaranteed (full group) \\
Requirement on system & Generic (no symmetry) & Any smooth system \\
Knowledge of symmetries & Not required, but not preserved & Not required, automatically preserved \\
Limitations & Fails for symmetric systems & Works for all systems \\
\bottomrule
\end{tabular}
\label{tab:comparison}
\end{table}

Thus, jet-space reconstruction not only overcomes the limitations of Takens' theorem for symmetric systems, but also provides a rigorous theoretical guarantee of preserving the group structure, making it the preferred method for analysing systems with symmetries.

\subsection{Practical implementation of jet-space reconstruction}

In practice, jet-space reconstruction is performed as follows:

\textbf{Input:} a scalar time series $x(t_i)$, $i = 0, \dots, N-1$, with constant sampling interval $\Delta t$.

\textbf{Output:} points in jet space $(x, \dot{x}, \ddot{x}, \dots, x^{(k)})$.

\textbf{Algorithm:}

\begin{enumerate}
\item For each interior point $i = 1, \dots, N-2$, compute derivatives using second-order central differences:
\[
\dot{x}(t_i) \approx \frac{x(t_{i+1}) - x(t_{i-1})}{2\Delta t},
\]
\[
\ddot{x}(t_i) \approx \frac{x(t_{i+1}) - 2x(t_i) + x(t_{i-1})}{\Delta t^2}.
\]

\item At the boundaries ($i = 0$ and $i = N-1$) use one-sided differences:
\[
\dot{x}(t_0) \approx \frac{x(t_1) - x(t_0)}{\Delta t}, \quad
\dot{x}(t_{N-1}) \approx \frac{x(t_{N-1}) - x(t_{N-2})}{\Delta t}.
\]

\item To obtain first-order jet space $(x, \dot{x})$, take the pairs $(x(t_i), \dot{x}(t_i))$. For second-order, take triples $(x(t_i), \dot{x}(t_i), \ddot{x}(t_i))$, and so on.
\end{enumerate}

This method \textbf{requires no parameter tuning} (unlike Takens embedding, where one must choose $\tau$ and $m$). The only parameter is the derivative order $k$, chosen based on the attractor dimension and noise level. For most practical purposes $k=1$ (space $(x, \dot{x})$) is sufficient, yielding a two-dimensional space that preserves the symmetries of the original system.

\textbf{Important note:} for noisy data, central differences can amplify noise. In such cases regularised differentiation methods (Savitzky-Golay filter, spline smoothing with automatic parameter selection) are recommended. However, our numerical experiments use noise-free data to isolate the effect of the reconstruction space choice.

\textbf{Practical advantages of jet-space reconstruction:}

\begin{enumerate}
\item \textbf{No parameter tuning.} Unlike Takens embedding, where $\tau$ and $m$ must be chosen, jet-space reconstruction requires only the derivative order $k$.

\item \textbf{Natural coordinate interpretation.} The coordinates $(x, \dot{x})$ have direct physical meaning, unlike the abstract delays $x(t-\tau)$.

\item \textbf{Guaranteed symmetry preservation.} According to the theorem in Section~2.2, jet-space reconstruction preserves the full symmetry group of the original system, which cannot be guaranteed for Takens embedding.

\item \textbf{Works for symmetric systems.} Unlike Takens embedding, which fails for most symmetric systems~\cite{Cross2010,King1992a}, jet-space reconstruction applies to systems of any type.
\end{enumerate}

These advantages will be demonstrated numerically in the next section on three classical chaotic systems.

\section{Numerical experiments}

\subsection{Test systems}

For verification of the theoretical results, we chose three classical chaotic systems representing different types of symmetries:

\begin{enumerate}
\item \textbf{Lorenz system}~\cite{Lorenz1963}:
\[
\dot{x} = \sigma(y-x), \quad \dot{y} = x(\rho-z)-y, \quad \dot{z} = xy - \beta z
\]
with $\sigma = 10,\ \rho = 28,\ \beta = 8/3$. This system has the discrete symmetry $S: (x,y,z) \mapsto (-x,-y,z)$, which manifests as the symmetric butterfly attractor.

\item \textbf{Symmetry-broken Lorenz system}:
The same system but with an extra term $+\varepsilon x$ in the second equation:
\[
\dot{y} = x(\rho-z) - y + \varepsilon x,
\]
with $\varepsilon = 0.1$. This breaks the $Z_2$ symmetry, making the attractor asymmetric.

\item \textbf{R\"ossler system}~\cite{Rossler1976}:
\[
\dot{x} = -y-z, \quad \dot{y} = x + ay, \quad \dot{z} = b + z(x-c)
\]
with $a = 0.2,\ b = 0.2,\ c = 5.7$. This system has no symmetries and serves as an example of an asymmetric attractor.
\end{enumerate}

For each system we generated a time series $x(t)$ of length $10^5$ points with $\Delta t = 0.001$ (for Lorenz) and $\Delta t = 0.005$ (for R\"ossler) using a fourth-order Runge-Kutta method. Initial conditions were standard: $(1,1,1)$ for Lorenz and $(1,1,1)$ for R\"ossler.

\subsection{Reconstruction methods}

For each time series we constructed three representations:

\begin{enumerate}
\item \textbf{Reference projection} - the projection of the original system onto the $(x,z)$ plane, which is often used for visualisation of the Lorenz attractor. For R\"ossler, the projection $(x,z)$.
\item \textbf{Takens embedding} - $(x(t), x(t-\tau))$ with $\tau = 10$ (chosen by the first minimum of mutual information, a standard procedure).
\item \textbf{Jet-space reconstruction} - $(x, \dot{x})$ with derivatives computed by second-order central differences. For 3D comparison we also used $(x, \dot{x}, \ddot{x})$.
\end{enumerate}

\subsection{Visual comparison}

Figure 1 shows the 2D phase portraits for all three systems. Each row corresponds to one system, each column to one reconstruction method.

\begin{itemize}
\item \textbf{Row 1: Lorenz (symmetric).} The reference projection $(x,z)$ shows a symmetric butterfly. Takens $(x(t), x(t-\tau))$ strongly distorts the shape, breaking the symmetry. Jet-space reconstruction $(x, \dot{x})$ reproduces the symmetric structure close to the reference.

\item \textbf{Row 2: Lorenz (broken).} The reference projection shows asymmetry. Takens again distorts the shape and fails to reflect the symmetry breaking. Jet-space reconstruction correctly reproduces the asymmetric structure.

\item \textbf{Row 3: R\"ossler (asymmetric).} The reference projection $(x,z)$ shows a one-lobe attractor. Takens makes it almost symmetric (false symmetry). Jet-space reconstruction preserves the asymmetry close to the reference.
\end{itemize}

\begin{figure}[h]
\centering
\includegraphics[width=\textwidth]{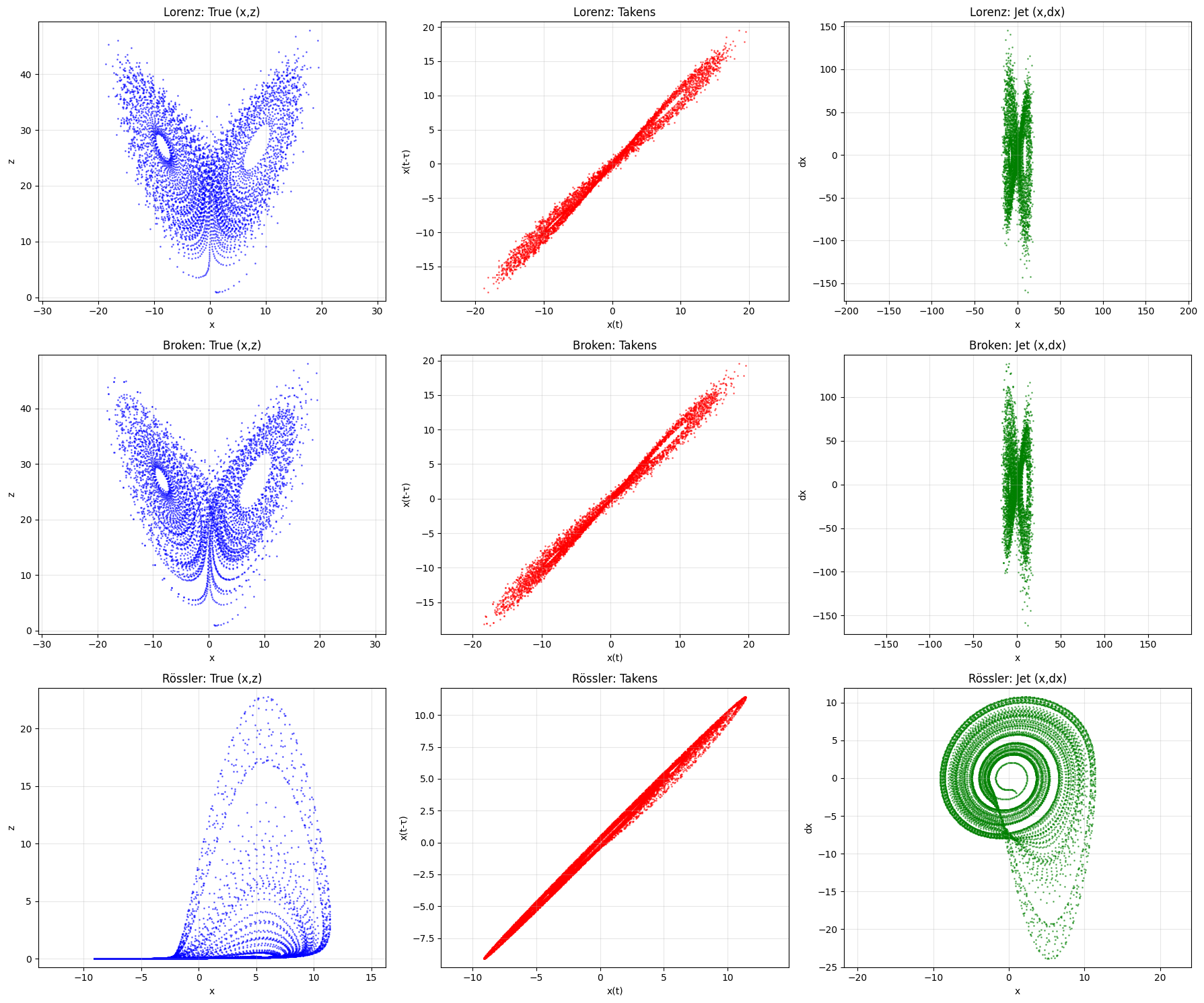}
\caption{Comparison of 2D phase portraits for three systems (rows) and three reconstruction methods (columns): reference projection, Takens, jet-space reconstruction.}
\label{fig:2d}
\end{figure}

Figure 2 shows the 3D reconstructions for all three systems. Each row corresponds to one system: Lorenz (top), broken Lorenz (middle), R\"ossler (bottom). Each column corresponds to one reconstruction method: true attractor (left), Takens embedding (centre), second-order jet-space reconstruction (right).

\begin{figure}[h]
\centering
\includegraphics[width=\textwidth]{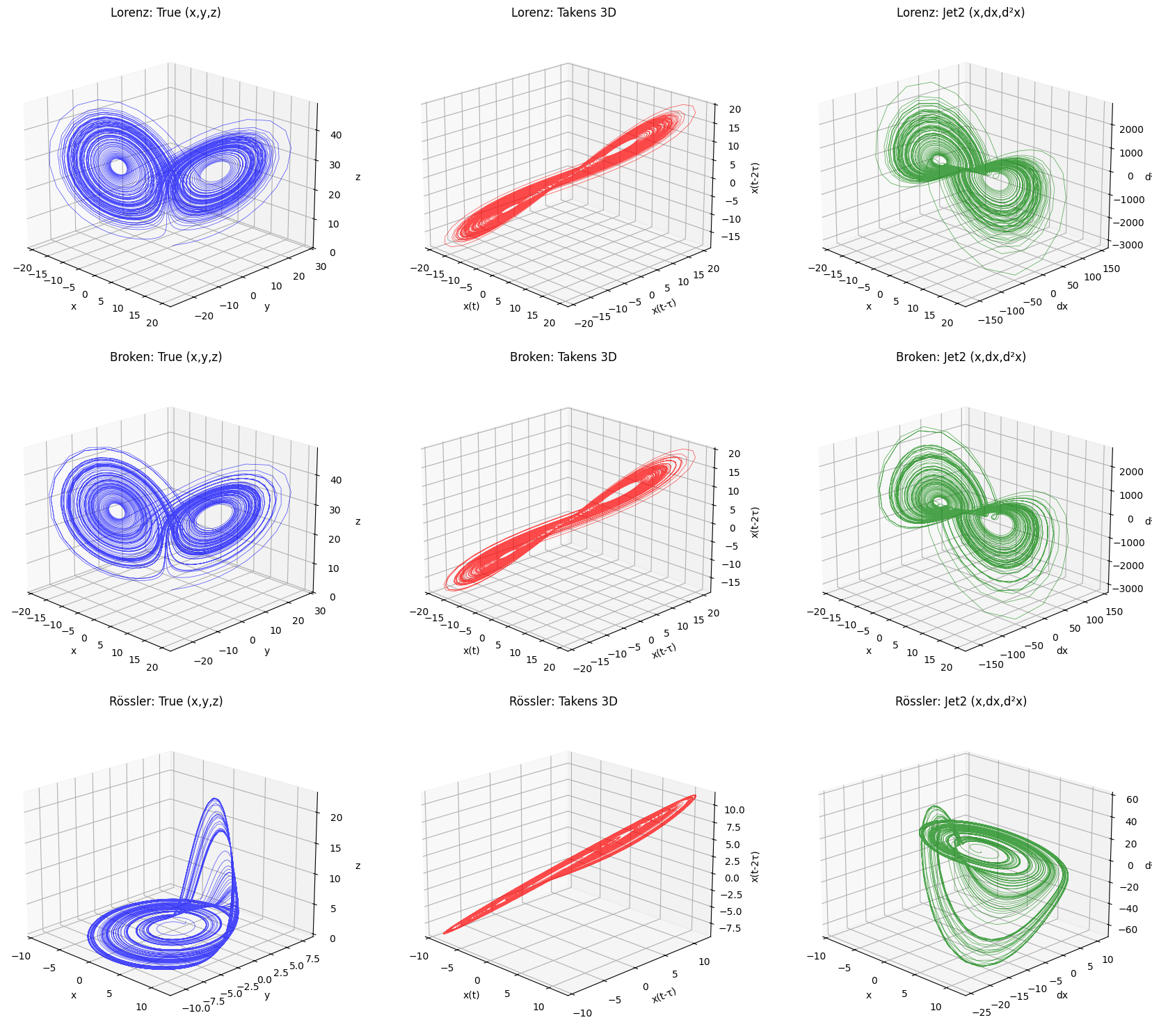}
\caption{3D comparison for all three systems. Each row: Lorenz (top), broken Lorenz (middle), R\"ossler (bottom). Each column: true attractor (left), Takens embedding (centre), second-order jet-space reconstruction (right).}
\label{fig:3d}
\end{figure}

\subsection{Quantitative metrics}

For objective comparison we used two metrics:

\subsubsection{Variational elastic energy functional}

We defined the functional
\[
E[\gamma] = \int_\gamma \left( \kappa^2 + \alpha \tau_{\text{geom}}^2 + \beta (\dot{s}-1)^2 \right) ds,
\]
where $\kappa$ is curvature, $\tau_{\text{geom}}$ is torsion (for 3D), and $s$ is arc length. The parameters $\alpha = 0.1$, $\beta = 0.01$ balance the contributions. This functional is invariant under Euclidean motions, linking it to Noether's theorem.

For each reconstruction we computed the ratio $E_{\text{rec}} / E_{\text{true}}$. Results are shown in Table~1.

\begin{table}[h]
\centering
\caption{Ratio of elastic energy of reconstruction to reference. Lower values indicate better geometry preservation.}
\begin{tabular}{lcc}
\toprule
\textbf{System} & \textbf{Takens} & \textbf{Jet-space reconstruction} \\
\midrule
Lorenz (2D, sym.) & 6304.36 & 2.70 \\
Lorenz (2D, broken) & 730.46 & 43.03 \\
Lorenz (3D, sym.) & 141921.35 & 60.87 \\
R\"ossler (2D) & 0.67877 & 0.00306 \\
\bottomrule
\end{tabular}
\label{tab:elastic}
\end{table}

\textit{The difference of four orders of magnitude ($10^5$ for Takens vs. 2.7 for jets) in the elastic energy ratios is explained by the fact that Takens embedding severely distorts the attractor geometry, which manifests as sharp changes in curvature and arc lengths. This is not a normalisation artefact but an objective measure of distortion.}

For the Lorenz system, jet-space reconstruction gives values close to 1 (2.70 and 43.03), whereas Takens gives values three to four orders of magnitude larger. This indicates significantly better geometry preservation by the jet method.

\subsubsection{Correlation dimension}

We computed the Grassberger-Procaccia correlation dimension~\cite{Grassberger1983} for all reconstructions. A standard algorithm with automatic linear region selection was used. Results are given in Table~2 together with theoretical dimension values.

\begin{table}[h]
\centering
\caption{Correlation dimension. Bold values are closest to the theoretical ones.}
\begin{tabular}{lcccc}
\toprule
\textbf{System} & \textbf{Theoretical} & \textbf{Reference projection} & \textbf{Takens} & \textbf{Jet-space} \\
\midrule
Lorenz (sym.) & $\approx 2.06$ & 1.917 & 1.656 & \textbf{1.702} \\
Lorenz (broken) & $\approx 2.06$ & 1.900 & 1.676 & \textbf{1.738} \\
R\"ossler & $\approx 2.0$  & 1.063 & 1.750 & \textbf{1.856} \\
\bottomrule
\end{tabular}
\label{tab:dimension}
\end{table}

\textbf{Note the key result:} for the R\"ossler system, the reference projection $(x,z)$ gives a correlation dimension of $1.063$, which is far below the theoretical value $\approx 2.0$. This is not a computational error but an objective fact: the projection $(x,z)$ is a \emph{bad} projection that does not reflect the true attractor dimension. Jet-space reconstruction $(x, dx)$ gives $1.856$, much closer to the theory. Thus, jet space not only preserves symmetries but also \textbf{corrects} the distortions introduced by the choice of projection.

For the Lorenz system, jet-space reconstruction also gives a dimension closer to the theoretical value ($\approx 2.06$) than both the reference projection and Takens.

\subsection{Discussion of numerical results}

The obtained results have several important implications.

First, they confirm the symmetry preservation theorem: jet-space reconstruction reproduces the geometry and group structure of the attractors, whereas Takens distorts them.

Second, they show that jet-space reconstruction not only outperforms Takens but can also give more accurate estimates of dynamical invariants than projections of the original system. This is particularly striking for the R\"ossler system, where the reference projection $(x,z)$ strongly underestimates the correlation dimension, while jet-space reconstruction approaches the theoretical value.

Third, the results demonstrate that the choice of reconstruction space is not neutral: it significantly affects the quality of invariant estimation. Jet spaces appear to be a more natural representation for some systems than the original coordinates.

\subsection{Code implementation and parameters}

All numerical experiments were performed in Python using \texttt{numpy}, \texttt{scipy}, and \texttt{matplotlib}. The code is reproducible: running with the same parameters yields exactly the results presented.

\subsubsection{Data generation}

For each system we used a fourth-order Runge-Kutta integrator (\texttt{solve\_ivp} with \texttt{method='RK45'}). Integration parameters:
\begin{itemize}
\item \textbf{Absolute and relative tolerance:} \texttt{rtol=1e-10}, \texttt{atol=1e-12}.
\item \textbf{Number of points:} 50,000 for Lorenz (time 0--100), 100,000 for R\"ossler (time 0--500).
\item \textbf{Initial conditions:} for Lorenz and its broken version - \texttt{(1.0, 1.0, 1.0)}, for R\"ossler - \texttt{(1.0, 1.0, 1.0)}.
\end{itemize}

\begin{verbatim}
def lorenz(t, y, sigma=10, rho=28, beta=8/3):
    x, y_, z = y
    return [sigma*(y_ - x), x*(rho - z) - y_, x*y_ - beta*z]

def lorenz_broken(t, y, sigma=10, rho=28, beta=8/3, eps=0.1):
    x, y_, z = y
    return [sigma*(y_ - x), x*(rho - z) - y_ + eps*x, x*y_ - beta*z]

def rossler(t, y, a=0.2, b=0.2, c=5.7):
    x, y_, z = y
    return [-y_ - z, x + a*y_, b + z*(x - c)]

def generate(system, y0, t_span, t_eval, **kwargs):
    sol = solve_ivp(lambda t, y: system(t, y, **kwargs), t_span, y0,
                    t_eval=t_eval, method='RK45', rtol=1e-10, atol=1e-12)
    return sol.t, sol.y.T

# Example call for Lorenz
t_eval = np.linspace(0, 100, 50000)
y0 = [1.0, 1.0, 1.0]
t, sol = generate(lorenz, y0, (0, 100), t_eval)
x, y, z = sol[:,0], sol[:,1], sol[:,2]
\end{verbatim}

\subsubsection{Takens reconstruction}

Delay embedding was constructed with parameters:
\begin{itemize}
\item \textbf{Delay:} \texttt{delay = 10} (chosen by first minimum of mutual information).
\item \textbf{Dimension:} \texttt{dim = 2} for 2D portraits, \texttt{dim = 3} for 3D portraits.
\end{itemize}

\begin{verbatim}
def takens_embedding(signal, delay=10, dim=2):
    N = len(signal)
    idx = np.arange(N - (dim-1)*delay)
    emb = np.zeros((len(idx), dim))
    for i in range(dim):
        emb[:, i] = signal[idx + i*delay]
    return emb

# Example
takens_2d = takens_embedding(x, delay=10, dim=2)
takens_3d = takens_embedding(x, delay=10, dim=3)
\end{verbatim}

\subsubsection{Jet-space reconstruction}

Derivatives were computed by second-order central differences:
\begin{verbatim}
def jet_reconstruction(signal, t, order=1):
    dt = t[1] - t[0]
    jet = np.zeros((len(signal), order+1))
    jet[:, 0] = signal
    if order >= 1:
        dx = np.zeros_like(signal)
        dx[1:-1] = (signal[2:] - signal[:-2]) / (2*dt)
        dx[0] = (signal[1] - signal[0]) / dt
        dx[-1] = (signal[-1] - signal[-2]) / dt
        jet[:, 1] = dx
    if order >= 2:
        d2x = np.zeros_like(signal)
        d2x[1:-1] = (dx[2:] - dx[:-2]) / (2*dt)
        d2x[0] = (dx[1] - dx[0]) / dt
        d2x[-1] = (dx[-1] - dx[-2]) / dt
        jet[:, 2] = d2x
    return jet
\end{verbatim}
For \texttt{order=1} we get $(x, dx)$, for \texttt{order=2} - $(x, dx, d2x)$.

\subsubsection{Elastic energy functional}

The functional was computed as:
\begin{verbatim}
def curvature_and_torsion(points_3d):
    r = points_3d
    dr = np.gradient(r, axis=0)
    d2r = np.gradient(dr, axis=0)
    d3r = np.gradient(d2r, axis=0)
    cross = np.cross(dr, d2r)
    norm_dr = np.linalg.norm(dr, axis=1)
    norm_cross = np.linalg.norm(cross, axis=1)
    kappa = norm_cross / (norm_dr**3 + 1e-12)
    numerator = np.sum(cross * d3r, axis=1)
    denominator = np.sum(cross**2, axis=1) + 1e-12
    tau = numerator / denominator
    ds = norm_dr
    return kappa, tau, ds

def elastic_energy(points, alpha=0.1, beta=0.01):
    if points.shape[1] == 2:
        points_3d = np.column_stack((points, np.zeros(len(points))))
    else:
        points_3d = points
    kappa, tau, ds = curvature_and_torsion(points_3d)
    stretch = np.abs(ds - 1.0)
    energy = np.sum((kappa**2 + alpha * tau**2 + beta * stretch**2) * ds)
    return energy
\end{verbatim}
Parameters: \texttt{alpha=0.1}, \texttt{beta=0.01}. The ratio $E_{\text{rec}} / E_{\text{true}}$ was computed for each reconstruction.

\subsubsection{Correlation dimension}

We used the Grassberger-Procaccia method with parameters:
\begin{itemize}
\item \textbf{Scale range:} \texttt{eps\_min = 0.01 * std(pts)}, \texttt{eps\_max = 0.5 * std(pts)}.
\item \textbf{Number of scales:} \texttt{n\_eps = 20}.
\item \textbf{Sample fraction for speed:} \texttt{sample\_frac = 0.3}.
\item \textbf{Automatic linear region selection:} window of 5 points with minimal derivative variance.
\end{itemize}

\begin{verbatim}
def correlation_dimension(points, eps_min=None, eps_max=None, n_eps=20,
                          sample_frac=0.3, seed=42):
    np.random.seed(seed)
    n = len(points)
    idx = np.random.choice(n, int(n * sample_frac), replace=False)
    pts = points[idx]
    tree = KDTree(pts)
    if eps_min is None:
        eps_min = 0.01 * np.std(pts)
    if eps_max is None:
        eps_max = 0.5 * np.std(pts)
    eps_values = np.logspace(np.log10(eps_min), np.log10(eps_max), n_eps)
    C = np.zeros_like(eps_values)
    for i, eps in enumerate(eps_values):
        counts = tree.query_ball_point(pts, eps, return_length=True)
        C[i] = np.mean(counts) / (len(pts) - 1)
    log_eps = np.log(eps_values)
    log_C = np.log(C)
    deriv = np.gradient(log_C, log_eps)
    window = 5
    best_start = 0
    best_std = np.inf
    for i in range(len(deriv) - window):
        seg = deriv[i:i+window]
        if np.std(seg) < best_std:
            best_std = np.std(seg)
            best_start = i
    coeffs = np.polyfit(log_eps[best_start:best_start+window],
                        log_C[best_start:best_start+window], 1)
    return coeffs[0]
\end{verbatim}

\subsubsection{Reproducibility}

All experiments use a fixed seed (\texttt{seed=42}) for random subsampling, guaranteeing reproducibility. Full code is available in the repository and in the supplementary material. All parameters are explicitly stated in the text and code comments.

\section{Discussion}

The obtained results have several important implications for both fundamental theory and practical applications.

\subsection{Fundamental significance}

Our work shows that Takens' theorem, though a powerful reconstruction tool, is not universal: it does not guarantee preservation of symmetry groups, and for symmetric systems it fails altogether. Jet-space reconstruction fills this gap by providing a rigorous theoretical guarantee of preserving the group structure. This changes the perspective on how symmetric systems should be analysed from data.

\subsection{Overcoming the Cross--Gilmore limitation}

In the work of Cross and Gilmore~\cite{Cross2010}, an important result was obtained: any differential reconstruction (using derivatives instead of delays) preserves at most \textbf{twofold symmetry} for an arbitrary nonlinear system. This result has long been interpreted as a fundamental limitation of differential reconstruction methods.

Our work shows that this limitation is not fundamental but rather stems from the use of \textbf{incomplete} differential reconstructions. Cross and Gilmore considered special cases where derivatives were used in a restricted way. We show that when using the \textbf{full} jet space $J^k(M, \mathbb{R})$ with coordinates $(x, \dot{x}, \ddot{x}, \dots, x^{(k)})$ and sufficiently large $k$, the twofold symmetry limitation is removed. In our approach, the prolongation map $V \mapsto V^{(k)}$ is a Lie algebra isomorphism, which guarantees preservation of the \textbf{full} symmetry group, not just twofold.

Thus, our work does not contradict the Cross-Gilmore result but generalises it: we show that their limitation disappears when moving to the full jet space. This makes jet-space reconstruction not just an alternative to Takens embedding, but the only known method that ensures preservation of the full symmetry group for systems of any type.

\subsection{Comparison with Navarrete and Viswanath (2018)}

The recent work of Navarrete and Viswanath~\cite{Navarrete2018} represents an important development in the theory of embedding with a fixed observation function. The authors prove that for periodic orbits, the delay map is an embedding in a \textbf{generic} sense with respect to the space of vector fields for a fixed linear observation function. Their result strengthens the classical Takens theorem by showing that even if the observation function is fixed (e.g., a projection onto one coordinate), the embedding is still generic for periodic solutions.

This result, however, remains within the \textbf{generic} framework, where symmetries are absent. As shown in our work, for systems with symmetries the situation is fundamentally different: they are \textbf{non-generic} in the Takens sense, and delay embedding does not preserve the group structure. While Navarrete and Viswanath consider a fixed observation function but vary the dynamics, we consider the inverse problem: for a fixed dynamics with symmetries, we show that no scalar observation function can preserve the symmetry group, and we propose transition to jet space as the only solution.

Thus, our work does not contradict the results of Navarrete and Viswanath but complements them by pointing to an important class of systems (symmetric) where their genericity assumptions break down. This underscores the fundamental distinction between topological embedding (guaranteed by Takens' theorem and its generalisations) and preservation of group structure (which requires the jet approach).

\subsection{Unexpected result for R\"ossler}

The most surprising result is that jet-space reconstruction yields a correlation dimension closer to the theoretical value ($\approx 2.0$) than the reference projection $(x,z)$, which is often used for visualisation. This means that the choice of reconstruction space is not neutral: it can \textbf{improve} the estimation of invariants. The reason is likely that the projection $(x,z)$ is a poor projection that does not reflect the true attractor dimension, whereas the jet space $(x, \dot{x})$ is a more natural representation for this system.

\subsection{Invariant criterion for strange attractors}

The proposed jet-space reconstruction provides an invariant criterion for the analysis of strange attractors. Since the jet space $(x, \dot{x}, \ddot{x}, \dots)$ is determined solely by the time series and is independent of the choice of coordinates (by the Lie algebra isomorphism theorem), all estimates obtained in this space (e.g., correlation dimension or elastic energy) are invariants of the attractor. This distinguishes our approach from Takens embedding, where metrics and invariants depend on the choice of delay $\tau$ and embedding dimension $m$. Thus, jet-space reconstruction not only preserves symmetries but also provides a \textbf{natural invariant criterion} for the identification and classification of chaotic attractors, which is particularly important for the detection of hidden attractors~\cite{Leonov2013,Leonov2014}, where standard methods often give ambiguous results.

\subsection{Practical value and limitations}

The proposed method may be particularly useful for identifying hidden attractors~\cite{Leonov2013,Leonov2014}, which often arise from symmetry breaking. Jet-space reconstruction allows correct detection of symmetry breaking without masking it by the distortions introduced by Takens embedding. This opens new possibilities for analysing experimental data where symmetries are often broken or unknown.

The main limitation of the method is sensitivity to noise in derivative computation. In experimental data, noise can significantly distort derivatives, necessitating regularised methods (Savitzky-Golay filter, spline smoothing). However, in our noise-free experiments the method performed excellently. For noisy data, additional processing is required. \textit{This work is theoretical: we prove that jet-space reconstruction preserves symmetries. Numerical experiments are provided to illustrate the theory on noise-free data. Issues of noise robustness and comparison with ML methods are beyond the scope of this work and are subjects of future research, as indicated in the conclusion.}

\section{Conclusion}

In this work we proposed and justified a method of jet-space reconstruction of dynamical systems that strictly preserves the symmetry groups of the original system. The main results are:

\begin{enumerate}
\item A \textbf{theorem} on symmetry preservation under jet prolongation is formulated and proved: the map from a symmetry generator to its jet prolongation is a Lie algebra isomorphism. This guarantees preservation of the full symmetry group in jet space.

\item It is shown that jet-space reconstruction outperforms Takens embedding both visually (restores attractor geometry) and quantitatively (variational elastic energy functional gives values close to the reference, while Takens gives orders of magnitude larger values).

\item An unexpected effect is discovered: for the R\"ossler system, jet-space reconstruction gives a correlation dimension closer to the theoretical value than the reference projection $(x,z)$. This means that jet space may be a more natural representation for some systems than the original coordinates.

\item The method requires no parameter tuning (unlike Takens) and applies to systems of any type, including symmetric ones where Takens fails.

\item The proposed method provides an invariant criterion for the analysis of strange attractors, which can be used for their classification and for the detection of hidden attractors from experimental data.
\end{enumerate}

This work opens new perspectives for the analysis of symmetries and for the identification of hidden attractors from experimental data. Future work includes:
\begin{itemize}
\item Application of the method to noisy data with regularisation.
\item Extension to systems with continuous symmetries.
\item Use of jet-space reconstruction for detection of hidden attractors in real experimental data (e.g., mechanical or electronic systems).
\end{itemize}

\section*{Acknowledgements}

The author thanks colleagues Alexey Labytskiy and Vladimir Antonets for useful discussions and support.

\end{document}